\documentclass[a4paper, reprint, showkeys, nofootinbib, twoside, pre, aps]{revtex4-1}
\usepackage[english]{babel}
\usepackage[utf8]{inputenc}
\usepackage{appendix}

\usepackage{amsfonts}
\usepackage{amsmath}
\usepackage{amssymb}
\usepackage[nolabel]{showlabels}

\usepackage{graphicx}
\usepackage[utf8]{inputenc}

\usepackage{xcolor}

\usepackage{tikz}
\usetikzlibrary{arrows.meta,backgrounds,positioning,fit, shapes.geometric}
\usetikzlibrary{shapes.misc}
\usepackage{forloop}
\usepackage{xintexpr}
\usepackage{calc}

\newcommand{\avg}[1]{\left\langle{#1}\right\rangle}
\newcommand{\qt}{{q_{\theta}}}

\newcommand{\dkl}{D_{KL}}
\newcommand{\ess}{\operatorname{ESS}}
\newcommand{\ket}[1]{#1\rangle}
\newcommand{\bra}[1]{\langle#1}
\newcommand{\tr}{\operatorname{Tr}}
\newcommand{\arctanh}{\operatorname{arctanh}}
\newcommand{\Js}{J_s}
\newcommand{\Jt}{J_\tau}
\newcommand{\Dt}{\Delta\tau}
\renewcommand{\v}[1]{\mathbf{#1}}
\newcommand{\bs}{\mathbf{s}}

\newcommand{\ssub}[1]{\v s\backslash \{{#1}\}}
\newcommand{\ssA}{\ssub{\mu_A,\nu_A}}

\tikzset{cross/.style={cross out, draw=black, minimum size=2*(#1-\pgflinewidth), inner sep=0pt, outer sep=0pt},
cross/.default={1pt}}

\begin{document}

\title{Estimation of the reduced density matrix and entanglement entropies using autoregressive networks}

\author{Piotr Białas}
\email{piotr.bialas@uj.edu.pl}
\affiliation{Institute of Applied Computer Science, Jagiellonian University, ul. \L ojasiewicza 11, 30-348 Krak\'ow, Poland}
\author{Piotr Korcyl}
\email{piotr.korcyl@uj.edu.pl}
\author{Tomasz Stebel}
\email{tomasz.stebel@uj.edu.pl}
\affiliation{Institute of Theoretical Physics, Jagiellonian University, ul. \L ojasiewicza 11, 30-348 Krak\'ow, Poland}
\author{Dawid Zapolski}
\email{dawid.zapolski@doctoral.uj.edu.pl}
\affiliation{Doctoral School of Exact and Natural Sciences, Jagiellonian University, Jagiellonian University, ul. \L ojasiewicza 11, 30-348 Krak\'ow, Poland}

\date{\today}

\begin{abstract}
We present an application of autoregressive neural networks to Monte Carlo simulations of quantum spin chains using the correspondence with classical two-dimensional spin systems. We use a hierarchy of neural networks capable of estimating conditional probabilities of consecutive spins to evaluate elements of reduced density matrices directly. Using the Ising chain as an example, we calculate the continuum limit of the ground state's von Neumann and Rényi bipartite entanglement entropies of an interval built of up to 5 spins. We demonstrate that our architecture is able to estimate all the needed matrix elements with just a single training for a fixed time discretization and lattice volume. Our method can be applied to other types of spin chains, possibly with defects, as well as to estimating entanglement entropies of thermal states at non-zero temperature.
\end{abstract}

\maketitle

\section{Introduction}

Entanglement entropy is a quantitative measure of quantum entanglement in a bipartite quantum system. It captures how much information is shared between two subsystems. Given a pure state $|\psi\rangle$ of a system, the density matrix is formally defined as
\begin{equation}
    \rho=|\ket{\psi}\bra{\psi}|.
\end{equation}
If we divide this system into subsystems $A$ and $B$, the von Neumann entanglement entropy of $A$ is defined as 
\begin{equation}
S(A) = -\tr \rho_A \log \rho_A,
\end{equation}
where  
\begin{equation}
\rho_A = \tr_B \rho    
\label{red_matr_def}
\end{equation}
is the reduced density matrix and $\tr_B$ is the partial trace over all degrees of freedom contained in part $B$.

It is often easier to compute the so-called Rényi entanglement entropy of order $n$, which is defined as: 
\begin{align}
S_n(A) = \frac{1}{1-n}\log \tr\, \rho_A^{n}.
    \label{Ren_entropy}
\end{align}
The von Neumann entropy can be recovered in the limit of $n \rightarrow 1$.

Entanglement entropy reflects the degree of quantum correlations between $A$ and $B$. If $A$ and $B$ are unentangled, $S(A) = 0$. The concept originated in early studies of black hole entropy and quantum fields in curved spacetime \cite{PhysRevD.34.373, PhysRevLett.71.666}.
Since then, it has become an important notion in quantum information theory and many-body physics. It provides insight into the structure of ground states in quantum systems, obeying area laws in gapped systems and logarithmic corrections in critical systems described by conformal field theory (CFT) \cite{2004JSMTE..06..002C, RevModPhys.82.277}. 

The efficiency of various methods of quantifying entanglement entropy depends on the nature of the system. Exact diagonalization of the Hamiltonian is limited to small system sizes. The density-matrix renormalization group method (DMRG) uses matrix product states to estimate the ground state \cite{2011AnPhy.326...96S} and is most effective in one dimension. The replica trick \cite{2004JSMTE..06..002C} employs the path integral formulation and often finds applications in Quantum Field Theory. In that case, entanglement entropy can be expressed as a ratio of partition functions that can be estimated using Quantum Monte Carlo methods \cite{PhysRevLett.104.157201,Humeniuk:2012xg}. Such a ratio can also be estimated using Jarzynski's theorem \cite{PhysRevLett.124.110602,Zhao_2022,Bulgarelli:2023ofi}. Finally, machine learning techniques are currently being applied to this problem as well \cite{Bialas:2024gha,Bulgarelli:2024yrz}.

In our previous work \cite{Bialas:2024gha}, we described a method of calculating Rényi entanglement entropies using generative neural networks. It was based on the replica trick and required calculating the partition function of the replica system. Indeed, neural networks with explicit probability estimation (e.g., autoregressive networks or normalizing flows) proved to be useful in the context of estimating partition functions \cite{2020PhRvE.101b3304N,PhysRevLett.126.032001}. However, the reformulation of the problem using the replica trick only allows for estimating Rényi entanglement entropies and does not give direct access to von Neumann entropy. 

In this work, we use autoregressive networks to explicitly estimate elements of the reduced density matrix, from which we can calculate eigenvalues and therefore evaluate many proposed measures of entanglement entropy. Since the size of the reduced density matrix grows exponentially with subsystem size, we limit ourselves to asymmetrical divisions where subsystem $A$ is much smaller than subsystem $B$. Our method relies on a trained hierarchy of autoregressive neural networks. The employed architecture is conditioned on the ket and bra defining the corresponding element of the reduced density matrix. We demonstrate that the architecture is able to learn the required conditional probabilities for all the matrix elements. Hence, our method allows us to calculate the full reduced density matrix with the neural network trained once.
Our approach is quite universal concerning boundary conditions and the form of the Hamiltonian.
We demonstrate it on the quantum Ising chain for which some analytical results exist, allowing for comparisons.

\begin{figure}
\centering
\begin{tikzpicture}[scale=0.85, every node/.style={scale=0.85}]
\newcounter{x}
\newcounter{y}
\newcounter{z}
\setcounter{z}{1}
\forloop{x}{0}{\value{x} < 8}{
    \forloop{y}{0}{\value{y} < 17}{

        \xintifboolexpr {(\value{y} == 0)}
        {
        \draw[line width=0.4mm, blue, dashed] (\value{x} - 0.5, \value{y}) -- (\value{x} - 0.2, \value{y});
        \draw[line width=0.4mm, blue, dashed] (\value{x} + 0.2, \value{y}) -- (\value{x} + 0.5, \value{y});
        \xintifboolexpr{ \value{x} < 3}{
        \setcounter{z}{\value{x} + 1}
        \node at (\value{x}, \value{y}) [circle, minimum size=0.8cm] {\arabic{z}};
        }
        {
        \setcounter{z}{\value{x} + 4}
        \node at (\value{x}, \value{y}) [diamond,minimum size=1cm] {\arabic{z}};
        }
        }
        {
        \xintifboolexpr {(\value{y} == 16)}
        {
        \draw[line width=0.4mm, blue, dashed] (\value{x} - 0.5, \value{y}) -- (\value{x} - 0.2, \value{y});
        \draw[line width=0.4mm, blue, dashed] (\value{x} + 0.2, \value{y}) -- (\value{x} + 0.5, \value{y});
        \xintifboolexpr{ \value{x} < 3}{
        \setcounter{z}{\value{x} + 4}
        \node at (\value{x}, \value{y}) [circle,minimum size=0.8cm] {\arabic{z}};
        }
        {
        \setcounter{z}{\value{x} + 4}
        \node at (\value{x}, \value{y}) [diamond,minimum size=1cm] {\arabic{z}};
        }
        }
        {
        \xintifboolexpr{\value{y} == 8}{
        \node at (\value{x}, \value{y}) [cross=0.2cm,black,draw] {};
        \node at (\value{x}, \value{y}) [cross=0.2cm,rotate=45,black,draw] {};
        }{
        \xintifboolexpr{\value{y} == 4 || \value{y} == 12 || \value{x} == 0 || \value{x} == 4}{
        \node at (\value{x}, \value{y}) [cross=0.2cm,black,draw] {};
        }{
        \xintifboolexpr{
        \value{x} /: 2 == 0 || \value{y} /: 2 == 0 
        }{
        \node at (\value{x}, \value{y}) [cross=3pt,black,rotate=45,draw] {};
        }{
        \filldraw[black] (\value{x}, \value{y}) circle (1.5pt);
        }
        }
    
        }
        }
        }
    }
}
\draw (-0.4, -0.4) rectangle (2.4, 0.4);
\node at (1, -0.6) {$\mu_A$};
\draw (2.6, -0.4) rectangle (7.4, 0.4);
\node at (5, -0.6) {$\mu_B$};
\draw (-0.4, 15.6) rectangle (2.4, 16.4);
\node at (1, 16.6) {$\nu_A$};
\draw (2.6, 15.6) rectangle (7.4, 16.4);
\node at (5, 16.6) {$\nu_B$};

\end{tikzpicture}
\caption{The structure of the spin configurations generated during reduced density matrix estimation, plotted for $L=8$, $m=16$ and $l=3$. Each node (number or mark) represents one classical spin (value -1 or 1). The first and last row are divided into two parts $(\nu_A,\nu_B)$ and $(\mu_A,\mu_B)$, those rows are sampled first, same numbers indicate identical spins. The rest of the configuration is filled in order: first: $\ast$, then: $\times$, then: $+$, and finally $\bullet$. The blue dashed lines mark connections with halved couplings.}
\label{fig:sample}
\end{figure}
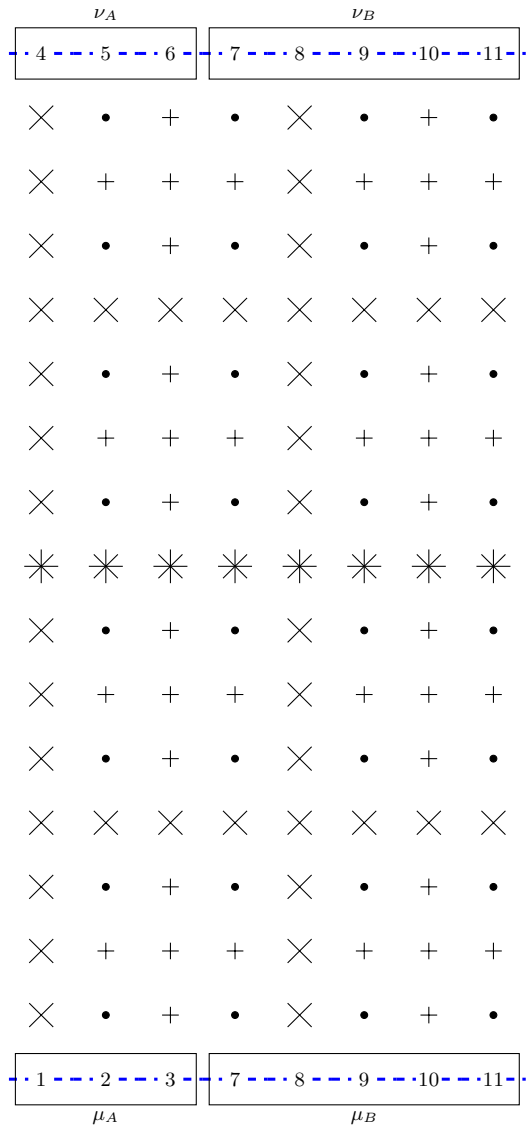

\section{Quantum Ising chain}

The Hamiltonian of 1D transverse Ising model with $L$ spins is given by:
\begin{equation}\label{eq:hamiltonian}
 H=-J\sum_{i=0}^{L-1} \hat\sigma^z_i\otimes \hat\sigma^z_{i+1} -h\sum_{i=0}^{L-1}\hat\sigma^x_i,
\end{equation}     
where $\hat\sigma^k_i$ is a $k$-th Pauli matrix associated to site $i$. We assume periodic boundary conditions.

For a system in thermal equilibrium at an inverse temperature $\beta=1/k_B T$, the density matrix is given by
\begin{equation}
\rho(\beta) =\frac{1}{Z}\sum_{i} e^{-\beta E_i} |\ket{\Phi^i}\bra{\Phi^i}|,
\end{equation}
where $|\ket{\Phi^i}$ are eigenstates of the Hamiltonian and 
\begin{equation}
  Z(\beta)  = \tr\rho(\beta) =\sum_{i} \bra{\Phi_i}|e^{-\beta H}|\ket{\Phi_i}
  \label{Z_def}
\end{equation}
is the partition function.

In what follows, we are interested in density matrix elements in the product states basis, $|\ket{\mu},|\ket{\nu} \in   \left\{|\ket{s_1}\otimes|\ket{s_2}\otimes \ldots |\ket{s_L} \right\}$:
\begin{equation}
  \rho_{\mu \nu}(\beta)  =\frac{1}{Z} \bra{\mu}|e^{-\beta H}|\ket{\nu}, 
  \label{rho_elements}
\end{equation}
where $|\ket{s_i}$ is an eigenvector of $\hat\sigma_i^z$ operator.

Applying path integral formalism we discretize the $[0,\beta]$ interval by dividing it into $m$ small intervals $\Dt$, $\beta=m \Dt$ \cite{PhysRevD.17.2637}. Then, the matrix elements of the 1D quantum Ising model can be calculated using the classical 2D Ising model with anisotropic couplings on the $L \times (m+1)$ lattice - see Figure \ref{fig:sample} for case $L=8$ and $m=16$. The numbers and marks on Figure \ref{fig:sample} denote classical (with values $-1$ or $1$) spins.

The energy of such a classical Ising model is given by:
\begin{equation}
\begin{split}
\tilde E(\bs) &= -\sum_{j=0}^{L-1} \Jt \sum_{i=0}^{m-1}   s_{i,j}\cdot s_{i+1,j}  \\
&\phantom{=}-\sum_{j=0}^{L-1}\Js \sum_{i=1}^{m-1}   s_{i,j}\cdot s_{i,j+1} \\
&\phantom{=}
-\frac{\Js}{2}\sum_{j=0}^{L-1}(s_{0,j}\cdot s_{0,j+1}+s_{m,j}\cdot s_{m,j+1}),
\end{split}
\label{energy_2D_ising}
\end{equation}
where
\begin{equation}\label{eq:ising-couplings}
\begin{split}
\Js\equiv \Js(J,h, \Dt)&=\Dt J,\\
\Jt \equiv \Jt(J,h, \Dt) & = \arctanh e^{-2 \Dt h}.
\end{split}
\end{equation}
Note that the last term of (\ref{energy_2D_ising}) is an artifact of the discretization we have chosen - the horizontal couplings between spins in the first and last row are divided by $1/2$ to avoid double counting.  In the horizontal direction of the 2D classical system, we have periodic boundary conditions since we are considering a periodic quantum spin chain. In the vertical direction, open boundary conditions are imposed (the first and last rows do not interact with each other).

To evaluate the density matrix (\ref{rho_elements}), we fix the spins in the first ($i=0$) and last ($i=m$) row of the spin configuration (see Figure \ref{fig:sample}). We perform the sum 
\begin{equation}
    [\rho^{cl}(m, \Dt)]_{\mu \nu}=\frac{1}{Z^{cl}(m, \Dt)} \sum_{\bs \backslash \{ \mu,\nu \}  }e^{-\tilde E(\bs) },
    \label{rho_class_Ising}
\end{equation}
where $\nu,\mu \in \{ 1, -1 \}^L$ denote the first and last row of the configuration, respectively. The sum is performed over all remaining spins in the 2D system. 
The partition function in (\ref{rho_class_Ising}) is given as
\begin{equation}
    Z^{cl}(m, \Dt)= \sum_{\bs|_{\mu=\nu}} e^{-\tilde E(\bs) },
    \label{Zmtau}
\end{equation}
where the first and last rows are the same, and the summation is performed over all spins in the configuration. Note that, due to the normalization, we have $\mathrm{Tr} \rho^{cl}(m, \Dt)= 1$, as required.

The quantum density matrix and partition function are obtained in the continuum limit,
\begin{equation}
\begin{split}
    [\rho^{cl}(m, \Dt)]_{\mu \nu} &\to \rho_{\mu \nu}(\beta), \\
    Z^{cl}(m, \Dt) &\to Z(\beta), 
\end{split}
    \label{Zlimit}
\end{equation}
when $m \to \infty$, $\Dt \to 0$ and $m \Dt = \beta=const$.
Note that in this limit, anisotropy increases, $\Js\rightarrow 0$, $\Jt\rightarrow \infty$.

We now divide the full spin chain of length $L$ into two segments, $A$ and $B$ of lengths $l$ and $L-l$, respectively. In Figure \ref{fig:sample}, the first and last rows of the 2D configuration were divided into parts $A$ and $B$, $\mu=(\mu_A,\mu_B)$, $\nu=(\nu_A,\nu_B)$.
To calculate $\rho_A$ we keep spins fixed in $\mu_A$ and $\nu_A$  and perform the summation over all possible configurations of spins in part $B$, where spins in $\mu_B$ are the same as in $\nu_B$ (in Figure \ref{fig:sample} same numbers indicate identical spins):
\begin{equation}
\begin{split}
  [\rho^{cl}_A(m, \Dt)]_{\mu_A \nu_A}  &=\sum_{\mu_B, \, \mu_B = \nu_B } [\rho^{cl}(m, \Dt)]_{\mu \nu}.
\end{split}  
  \label{red_dens_sum}
\end{equation}
The last equation can also be written using  (\ref{rho_class_Ising}):
\begin{equation}
  [\rho^{cl}_A(m, \Dt)]_{\mu_A \nu_A}=  Z^{cl}(m,\Dt)^{-1} \sum_{\bs \backslash \{ \mu_A,\nu_A \}  }e^{-\tilde E(\bs) },
  \label{red_dens_sum_rewritten}
\end{equation}
where by $\bs \backslash \{ \mu_A,\nu_A \}$ we denote all the spins in the classical configuration which are not part of $\mu_A$ or $\nu_A$.
The reduced density matrix (\ref{red_matr_def}) is obtained from (\ref{red_dens_sum_rewritten}) by taking the continuum limit.

The sum $\sum_{\bs \backslash \{ \mu_A,\nu_A \}  }e^{-\tilde E(\bs) }$ appearing in (\ref{red_dens_sum_rewritten}) we shall for simplicity call the partition function and denote by $Z^{cl}_{ \mu_A,\nu_A}(m,\Dt)$. Except for very small system sizes,  such a sum cannot be performed in practice due to the large number of classical spin configurations. In this manuscript, we propose to use autoregressive neural networks to estimate partition functions $Z^{cl}_{ \mu_A,\nu_A}(m,\Dt)$. We demonstrate that such partition functions can be estimated using one model of autoregressive generative neural networks for all $\mu_A$ and $\nu_A$.

\section{Method}

\subsection{Autoregressive neural networks}

Given a configuration of spins $\v s$, autoregressive neural network (ANN) \cite{2015arXiv150203509G,2019PhRvL.122h0602W}  can provide the probability $\qt(\v s)$
which is 
factorized into a product of conditional probabilities  
\begin{equation}
\label{eq:factorisation}
     q_{\theta}(\v s)  =  q(s^1)\prod_{i=2}^{N} q_{\theta}(s^i|s^1,s^2,\dots,s^{i-1}),
 \end{equation} 
 where $N$ is the number of spins. 
 Given this decomposition, we can generate a sample using {\em ancestral sampling}. We start by generating $s^1$ with probability $q(s^1)$. This allows us to calculate the conditional probability $q(s^2|s^1)$ and hence generate spin $s^2$, which in turn allows us to calculate $q(s^3|s^1,s^2)$.  We continue in this way until all the spins are generated.

The network is trained so that $\qt$ is as close as possible to the target distribution, which in our case is given by the Boltzmann probability distribution,
\begin{equation}
p(\v s) = \frac{1}{Z} e^{-\tilde E(\v s)}.
\label{eq. p}
\end{equation}
The training is done by minimizing the difference between $\qt$ and $p$ as measured by 
the Kullback-Leibler divergence:
\begin{align}
    \dkl (q_\theta | p) &= \sum_{\v s} q_\theta(\v s) \, \log \left(\frac{q_\theta(\v s)}{p(\v s)}\right).
    \label{eq:KL_loss} 
\end{align}

The decomposition of $\qt(\v s)$ into the product \eqref{eq:factorisation} can be done in an arbitrary order. In the original approach, called Variational Autoregressive Network (VAN) \cite{2019PhRvL.122h0602W}, the generation of spin configuration was performed row by row. 
In \cite{Bialas:2022qbs}, we took advantage of this freedom as well as the Markov property of the Ising model, and proposed to generate spins hierarchically, using several neural networks instead of one. This approach, called Hierarchical Autoregressive Networks (HAN), provides faster training and considerably improves its efficiency. In what follows, we shall use HAN to sample configurations needed for the calculation of sum (\ref{red_dens_sum_rewritten}). 
The architecture we propose in this work differs from the original HAN architecture by the addition of a row of spins (see Figure~\ref{fig:sample}) and the order of sampling the spins. There is still some freedom of ordering the spins left in HAN and we use it to order the spins included in $\mu_A$ and $\nu_A$ at the beginning: $\mu_A=\{s^1,\ldots,s^{l}\}$, $\nu_A=s^{l+1},\ldots s^{2l}$ (denoted by numbers 1-6 in Figure \ref{fig:sample}). 
If we fix those spins, we can train the neural network to approximate the conditional probability
\begin{equation}
    q(\v s  \backslash \{ \mu_A,\nu_A \} )\equiv \prod_{i=2l+1}^{N} q(s^i|s^1,\ldots,s^{2l},\ldots, s^{i-1}).
\end{equation}
In practice, we train the network on all the combinations of spins $s^1,\ldots,s^{2l}$ by drawing them at random from a uniform distribution (see Section~\ref{sec:han-training}).

\subsection{Neural Importance Sampling (NIS)} 
   
For the estimation of partition functions
\begin{equation}
    Z^{cl}_{ \mu_A,\nu_A}(m,\Dt)= \sum_{\ssA  }e^{-\tilde E(\bs) },
\end{equation}
one can use Neural Importance Sampling (NIS) \cite{2020PhRvE.101b3304N}. To this end, we rewrite:
\begin{equation}
\begin{split}
        Z^{cl}&_{ \mu_A,\nu_A}(m,\Dt)=\\
        &\sum_{\ssA} \qt(\ssA) 
        \hat w_A(\v s) \\
       &\equiv \avg{\hat w_A(\v s)}_{q_{\theta}},
    \label{NIS_for Z} 
\end{split}    
\end{equation}
where we defined importance weights,
\begin{equation}
   \hat w_A(\v s) =  e^{-\tilde E(\v s)}/ \qt(\bs \backslash \{ \mu_A,\nu_A \} ).
\end{equation}
Equation (\ref{NIS_for Z}) can be treated as a standard Monte Carlo average; therefore, we can estimate it using some relatively small number (compared to the number of states) of samples:
\begin{equation}
\begin{split}
      Z^{cl}_{ \mu_A,\nu_A}(m,\Dt) &\approx \frac{1}{N_{s}}  \sum_{i=1}^{N_{s}} \hat w_A(\v{s}_i),\\ 
      \v s_i\backslash \{ \mu_A,\nu_A \} &\sim \qt(\bs \backslash \{ \mu_A,\nu_A \} ),
\end{split}      
 \label{MC_average_Z}
\end{equation}
where we denoted that the part of the configuration which is not $\mu_A$ or $\nu_A$ is drawn from the distribution $\qt(\bs \backslash \{ \mu_A,\nu_A \} )$. 

\subsection{Sampling}

The sampling from the conditional probability distribution $\qt(\ssA)$ proceeds as follows:
\begin{enumerate}
\item Given two product states of subsystem $A$, $|\ket {\mu_A}$ and $|\ket{\nu_A}$, we fix the classical spins in the first and last row (denoted as numbers 1 to 6 in Figure \ref{fig:sample}). 
\item We draw conditionally the spins from part $B$ (numbers 7-11 in Figure \ref{fig:sample}) using the first autoregressive network from HAN. The probability of spins in part $B$ depends on spins $A$. Spins from part $B$ are copied to the first and last row of the configuration. 
\item  Once the upper and lower rows are fixed, the interior of the configuration is filled. We make use of the Hammersley-Clifford theorem \cite{Hammersley-Clifford, Clifford90markovrandom}, which assures that once a closed loop of spins is fixed, the probability of loop-interior spins does not depend on the exterior. This is the key observation for the HAN algorithm \cite{Bialas:2022qbs}, which is applied here. We use a hierarchy of networks to fix groups of spins; in Figure \ref{fig:sample} we schematically show such a division with different marks.

We choose the vertical size $m$ (where, $m=\beta/\Dt$) of the configuration to be a multiple of $L$: $m=kL, \ k=2,3,4,\ldots$.
To form loops of spins, we divide the interior of the configuration into squares of size $L\times L$ by fixing spins on horizontal lines (denoted by~"$\ast$" in Figure \ref{fig:sample}) - this is done iteratively: if $k$ is even, we split the configuration in half by a line of spins, and if $k$ is odd, we effectively reduce it by one "cutting off" one square of size $L\times L$ using the line of spins. This procedure is continued recursively until only squares remain.

Fixing the spins is done with further networks from the hierarchy, which sample spins using probabilities depending on the spins fixed previously.

\item At the next step, $\times$-spins are drawn using yet another network and they depend only on their surrounding spins (16 in total for $L=8$). 
Since the functional dependencies on the outside spins are the same for each remaining square of the configuration, we can use the same network, run in parallel, to fix "$\times$"-spins in each square. The last network from the hierarchy can be used to fix "$+$"-spins. The remaining "$\bullet$"-spins are drawn using the heatbath algorithm. This hierarchy is extended for the case of $L$ larger than 8. 
\end{enumerate}

\subsection{Training of HAN networks} 
\label{sec:han-training}

We follow the original VAN/HAN approach and apply backward training, namely, the samples for training are provided by the network itself. Since we calculate partition functions (\ref{MC_average_Z}) for each of the states $\mu_A$ and $\nu_A$ independently, we could, in principle, train a separate HAN for each of the matrix elements. This would, however, increase the numerical cost of the whole simulation, so instead, we train one HAN for all the matrix elements. To be specific, the batch of samples for training is obtained by: {\it i)} drawing (with probability $1/2$) all spins for states $\mu_A$ and $\nu_A$, {\it ii)} sampling other spins from $\qt(\bs \backslash \{ \mu_A,\nu_A \} )$ according to the algorithm explained in the previous section. 

The training is performed by minimizing the $D_{KL}$--based loss function, which is calculated using a batch of configurations sampled from $\qt(\bs \backslash \{ \mu_A,\nu_A \} )$ as:
\begin{equation}
F_q= \frac{1}{N_{batch}}\sum_{k=1}^{N_{batch}} \left[\tilde E(\mathbf{s}_k)+\log \qt(\bs_k \backslash \{ \mu_A,\nu_A \} ) \right].
\label{F_q_def}
\end{equation}
Since the batch contains configurations with different states $\mu_A$ and $\nu_A$, the network is trained to generate configurations for all reduced density matrix elements (\ref{red_dens_sum_rewritten}).

Some additional details on the neural network architecture and training quality are described in Appendix \ref{app}.

\section{Results}

In the previous section, we described the method of sampling spin configurations, which gives the density matrix at finite $\Delta\tau$ for a given temperature $\beta$ (given $m$). In this section, we shall present numerical results for the quantum Ising model (\ref{eq:hamiltonian}). We choose $J=h=1$, so that the system is at the critical point. All the results we shall present are obtained for a spin chain of size $L=32$ spins.

For simplicity, we consider only several values of $m$: $m=kL$ and $k=2,3,\ldots,8$, which are then extrapolated to $k=\infty$. We evaluate the density matrix elements at 5 linearly spaced values of $\Dt$: $\Dt=0.4, 0.35, 0.3, 0.25, 0.2$, which allows us to make an extrapolation $\Dt \to 0$. For each choice of parameters $(l,m,\Dt)$, we train a separate hierarchy of networks.


\subsection{Reduced density matrix}

\begin{figure}
    \centering
    \includegraphics[width=1.\linewidth]{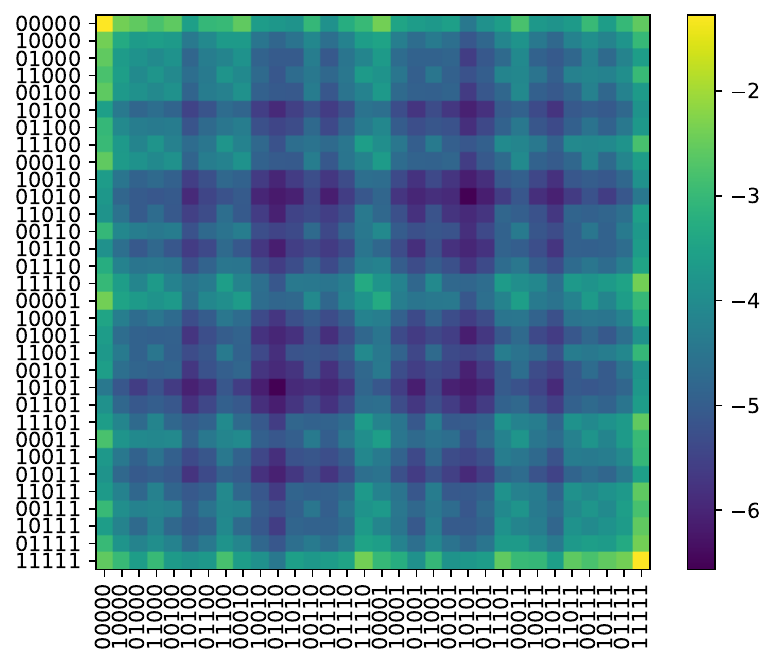}
    \caption{Natural logarithm of the density matrix elements in the product base $\left\{|\ket{s_1}\otimes|\ket{s_2}\otimes \ldots |\ket{s_5} \right\}$. Plotted for $L=32$, $l=5$, $k=8$, $\Dt = 0.2$. On the axes we introduced notation: $1$: spin has value $+1$, $0$: spin has value $-1$. }
    \label{fig:matrix}
\end{figure}

We start with calculating the reduced density matrix (\ref{red_dens_sum_rewritten}) for $k=8$ and $\Dt=0.2$. The classical spin configurations (for $L=32$) are of size $32\times 256$ spins ($k=8$). We choose the subsystem $A$ to be of size $l=5$, for which the reduced density matrix is of size $32\times32$. In Figure \ref{fig:matrix} we show the result of the estimation of density matrix elements. On the horizontal and vertical axes, we denoted $32$ product states; for clarity, we introduced notation: $1$ - spin has value $1$, $0$ - spin has value $-1$. To better visualize the order of magnitude of matrix elements, we plotted the logarithm of their values (this is possible as the values of the reduced density matrix elements are real and positive in this representation).%

We first note that the density matrix is symmetric w.r.t. the diagonal and antidiagonal, which reflects the hermiticity of $\rho_A$ and $Z_2$ symmetry of the Ising model and the specific choice of our basis of the Hilbert space. Looking at the values of the matrix elements, one sees that the biggest values correspond to elements 
\begin{eqnarray}
&\bra{1,1,1,1,1}|\rho_A|\ket{1,1,1,1,1},\nonumber \\ &\bra{-1,-1,-1,-1,-1}|\rho_A|\ket{-1,-1,-1,-1,-1}, \nonumber 
\end{eqnarray}
 and the smallest to 
 \begin{eqnarray}
\bra{-1,1,-1,1,-1}&|\rho_A|&\ket{1,-1,1,-1,1},\nonumber \\ \bra{1,-1,1,-1,1}&|\rho_A|&\ket{-1,1,-1,1,-1}. \nonumber 
\end{eqnarray}

\begin{figure}
    \centering
    \includegraphics[width=1.05\linewidth]{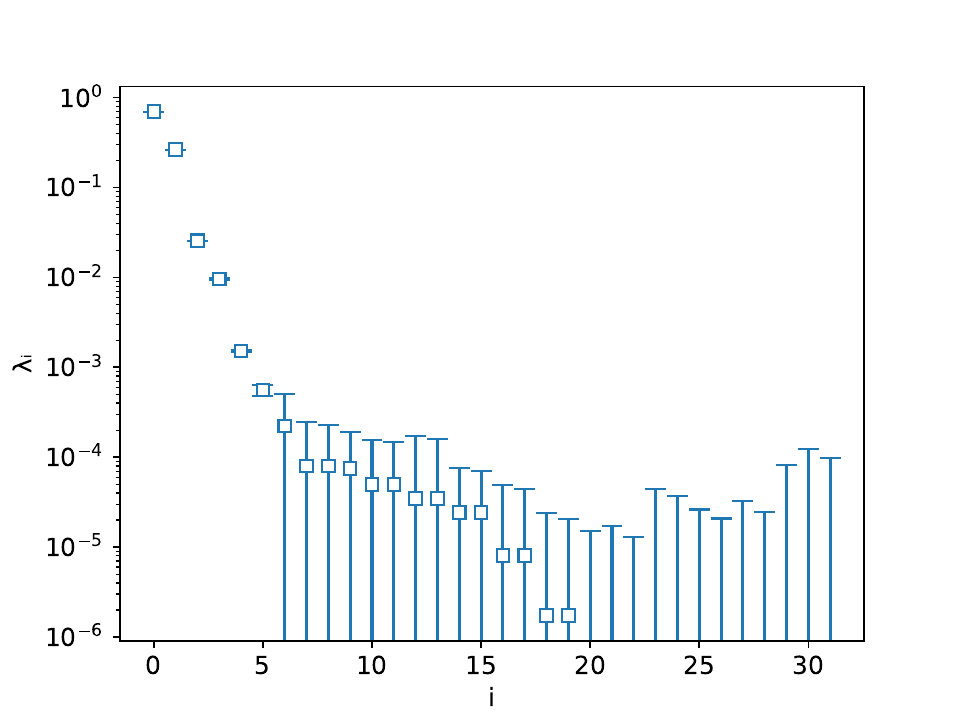}
    \caption{Eigenvalues of reduced density matrix $\{\lambda_i \}_{i=0,\ldots,2^l-1}$ obtained for $L=32$, $l=5$, $k=8$, $\Dt=0.2$. They were plotted in terms of ordinal numbers (from biggest to smallest).}
    \label{fig:eigenvalues_plot}
\end{figure}

The density matrix elements themselves depend on the basis of states; on the other hand, full physical information is encoded in the eigenvalues of $\rho$. In Figure \ref{fig:eigenvalues_plot}, we plot all 32 eigenvalues of the reduced density matrix discussed above. We obtained them by direct diagonalization of the density matrix using the \texttt{numpy} package. The eigenvalues were sorted and plotted in terms of their ordinal number. Errors were obtained using the bootstrap method with 800 samples. We see that only the 6 biggest eigenvalues are nonzero within the errors. The values decrease roughly exponentially for the biggest values. The first eigenvalue is of order 1, whereas the 6th one is already of the order $10^{-3}$.  

\subsection{Entanglement Entropies}

In this section, we present results for von Neumann and Rényi entropies. They are obtained from eigenvalues of the reduced density matrix $\{\lambda_i \}_{i=0,\ldots,2^l-1}$ as:
\begin{align}
    S(A) &= \sum_{i=0}^{2^l-1} \lambda_i \log \lambda_i, \\
    S_n(A) &= \frac{1}{1-n}\log \sum_{i=0}^{2^l-1} \lambda_i^n.
    \label{EE_from_lambda}
\end{align}
The uncertainties of entanglement entropies were obtained using the bootstrap method.

Results presented in this section are obtained after taking the zero temperature limit, $\beta \to \infty$, in order to study the entanglement entropy of the ground state. 
In this case, we perform the limit by first taking $k\to \infty$ and then $\Dt\to 0$. Below we present the details of this procedure.

\subsubsection{Zero temperature and continuum extrapolations}

\begin{figure}
    \centering
    \includegraphics[width=1.05\linewidth]{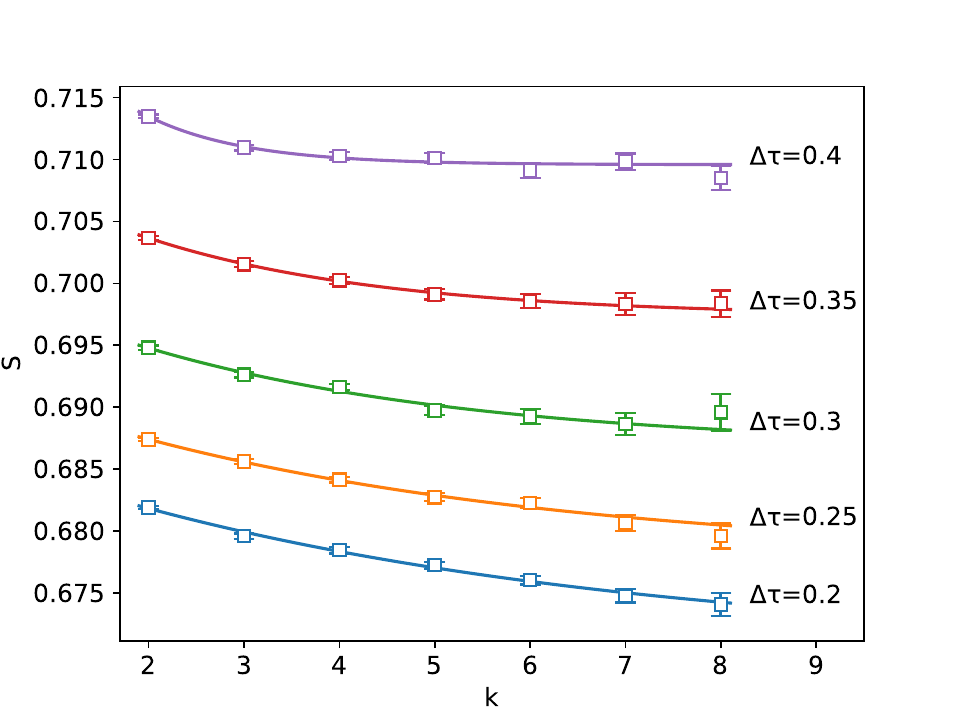}
    \caption{Von Neuman entanglement entropy as a function of size of the temperature dimension, $k=\beta/(\Dt L)$, plotted for several values of discretization parameter $\Dt$: 0.2, 0.25, 0.3, 0.35, 0.4 (denoted by points with different colors); $L=32$ and $l=3$. For each $\Dt$ we fit the function $y_{\Dt_i}(k)$, see Eq.~(\ref{combined_fit}), which we denote by curves. }
    \label{fig:vN_dt_k_fixed}
\end{figure}

In Figure \ref{fig:vN_dt_k_fixed} we show von Neumann entanglement entropy as functions of $k$ for several values of $\Dt$ (points plotted with different colors). The corrections coming from finite $k$ decay exponentially with $k$ \cite{Bialas:2024gha} hence the extrapolation $k \to \infty$ is performed by fitting a function of the form $y_{\Dt}(k)=a_{\Dt}+b_{\Dt}\, \exp(-c_{\Dt} k)$ (curves in Figure \ref{fig:vN_dt_k_fixed}). For each $\Dt$ we therefore extract the value of entanglement entropy at $k=\infty$, which is the parameter $a_{\Dt}$ in our fits.

\begin{figure}
    \centering
    \includegraphics[width=1.05\linewidth]{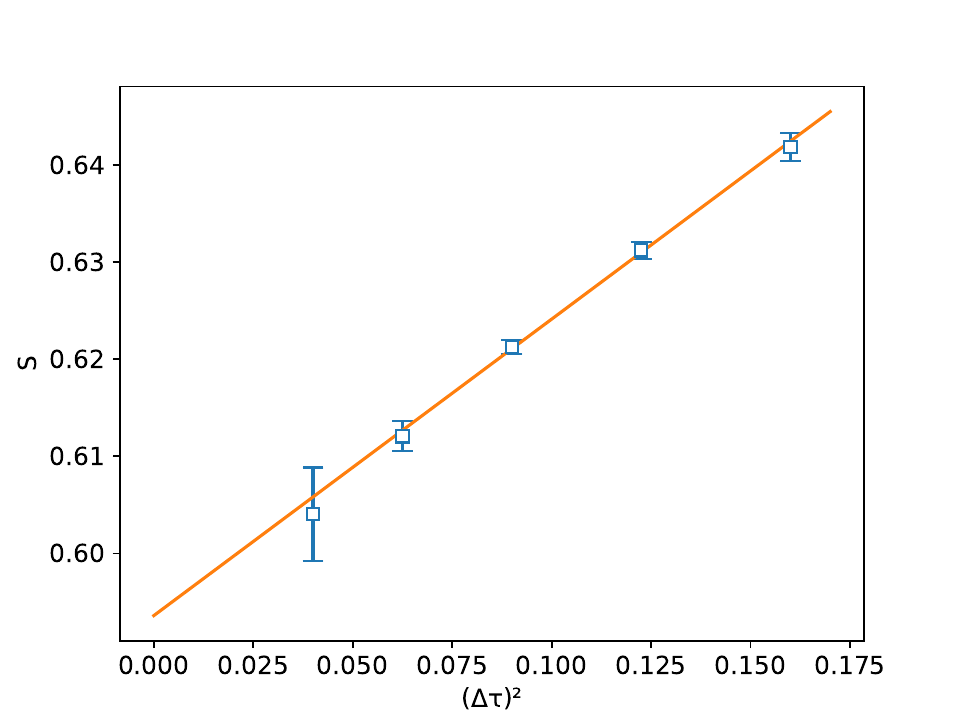}
    \caption{Von Neumann entropy extrapolated to $k = \infty$ as a function of $(\Dt)^2$ for $l=2$.}
    \label{fig:S(dt^2)}
\end{figure}

In Figure \ref{fig:S(dt^2)} we show fitted values of $y_{\Dt}(k=\infty)=a_{\Dt}$ as a function of the square of parameter $\Dt$. Since the behavior we observe seems to be linear, we postulate the $\sim(\Dt)^2$ behavior of the leading corrections coming from the finite discretization constant $\Dt$. 

To reduce the uncertainty of obtained results for the entropy at $k=\infty$ and $\Dt=0$, we use a combined fit in $k$ for all $\Dt$ where $a_{\Dt}$ is quadratic in $\Dt$. To be more specific, we employ the following parametrization of data for entanglement entropy:
\begin{equation}
    \begin{cases}
        y_{\Dt_1}(k) = S+ a_1 (\Dt_1)^2 + b_{\Dt_1}\, \exp(-c_{\Dt_1} k)\\
        y_{\Dt_2}(k) = S+ a_1 (\Dt_2)^2 + b_{\Dt_2}\, \exp(-c_{\Dt_2} k)\\
        \cdots \\
        y_{\Dt_p}(k) = S+ a_1 (\Dt_p)^2 + b_{\Dt_p}\, \exp(-c_{\Dt_p} k)\\
    \end{cases}
    \label{combined_fit}
\end{equation}
where $S$ corresponds to the entanglement entropy in the zero temperature and continuum limit, and $p$ is the number of values of $\Dt$ (in our case $p=5$).

To assess the systematic uncertainty of the extrapolated value, we also consider including terms proportional to $(\Dt)^3$ in (\ref{combined_fit}) and perform a second fit\footnote{We have also checked the terms $\sim(\Dt)^4$, and the error is similar.}. The difference between the values of $S$ obtained with the two functional forms is treated as the systematic uncertainty of our result, which is added in quadrature to the statistical uncertainty.

\subsubsection{Ground state entanglement }

\begin{figure}
    \centering
    \includegraphics[width=1.05\linewidth]{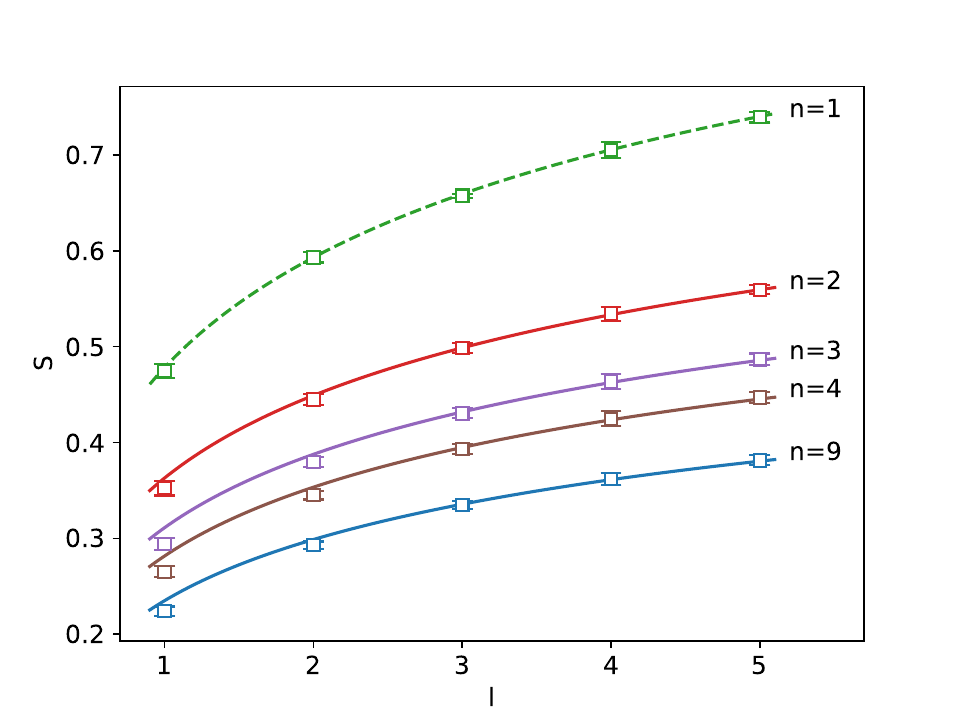}
    \caption{von Neumann ($n=1$) and Rényi entropies ($n\ge 2$)  as functions of subsystem size $l$ for spin chain with $L=32$ spins at the zero temperature (points). 
    The dashed line corresponds to the conformal field theory prediction \cite{2004JSP...116...79J}, see Eq.~(\ref{eq_CFT_vNE}). Solid lines represent fits of equation (\ref{eq_CFT_RE}) to our results for $3\le l \le 5$. }
    \label{fig:entr_system_size}
\end{figure}

We present the von Neumann and Rényi entanglement entropies as functions of $l$, the size of the subsystem $A$. We keep the size of the whole system as $L=32$. One expects that $\lim_{n \to 1} S_n(A)=S(A)$; therefore, we denote a value of the von Neumann entropy as $n=1$.
In Figure \ref{fig:entr_system_size} we show $S(l)$ and $S_n(l)$ for $n=2,3,4,9$ (points) - note that we skip some values of $n$ for the brevity of the plot. As we noted before, we limit ourselves to $l\le 5$, so that the reduced density matrices are relatively small. The uncertainties are at the level of one percent or less. 

We shall now compare our results with the prediction of CFT. The von Neumann entropy is obtained as \cite{PhysRevLett.90.227902,Calabrese_2009}:
\begin{equation}
    S^{CFT}(l)= \frac{1}{6} \ln\left[ \frac{L}{\pi}\sin\left( \frac{\pi l}{L} \right) \right] + b_1, 
    \label{eq_CFT_vNE}
\end{equation}
where the constant coefficient $b_1\approx 0.478558$ is known with numerical precision \cite{2004JSP...116...79J,Cardy:2007mb, PhysRevB.111.104437}. In Figure \ref{fig:entr_system_size}, we plotted the CFT result (\ref{eq_CFT_vNE}) using a dashed curve.  The agreement is very good - the $\chi^2$ per degree of freedom ($\chi^2/d.o.f.$) is 0.3. This comparison serves as a cross-check of our results.

Rényi entropy in the CFT is given by \cite{Calabrese_2009}:
\begin{equation}
    S^{CFT}_n(l)= \frac{1}{12}\left( 1+\frac{1}{n} \right) \ln\left[ \frac{L}{\pi}\sin\left( \frac{\pi l}{L} \right) \right] + b_n, 
    \label{eq_CFT_RE}
\end{equation}
where $b_n$ are constants depending only on $n$ when $1\ll l\ll L$. To our knowledge, for Rényi entropies, $b_n$ are unknown, hence, we cannot make a direct comparison with our results. However, we perform fits to our data of equation (\ref{eq_CFT_RE}) for each $n\ge 2$ separately with only one free parameter, $b_n$. Taking all values, at $l=1,2,3,4,5$, we get  $\chi^2/d.o.f.$ between 0.5  and 1.7, which means that our data are not exactly described by (\ref{eq_CFT_RE}). This is not a surprise \cite{2009JSMTE..02..063N,Cardy:2010zs,Chanda:2023zcg} as the condition $1\ll l$ is not fulfilled. Taking values for $l=3,4,5$ already leads to very low $\chi^2/d.o.f.$ between 0.01 and 0.09. In Figure \ref{fig:entr_system_size}  we showed such curves obtained with the fit to only 3 points - this does not provide strong statistical evidence that (\ref{eq_CFT_RE}) is fulfilled, 
 especially because we expect large corrections due to the smallness of $l$.

 \begin{figure}
    \centering
    \includegraphics[width=1.05\linewidth]{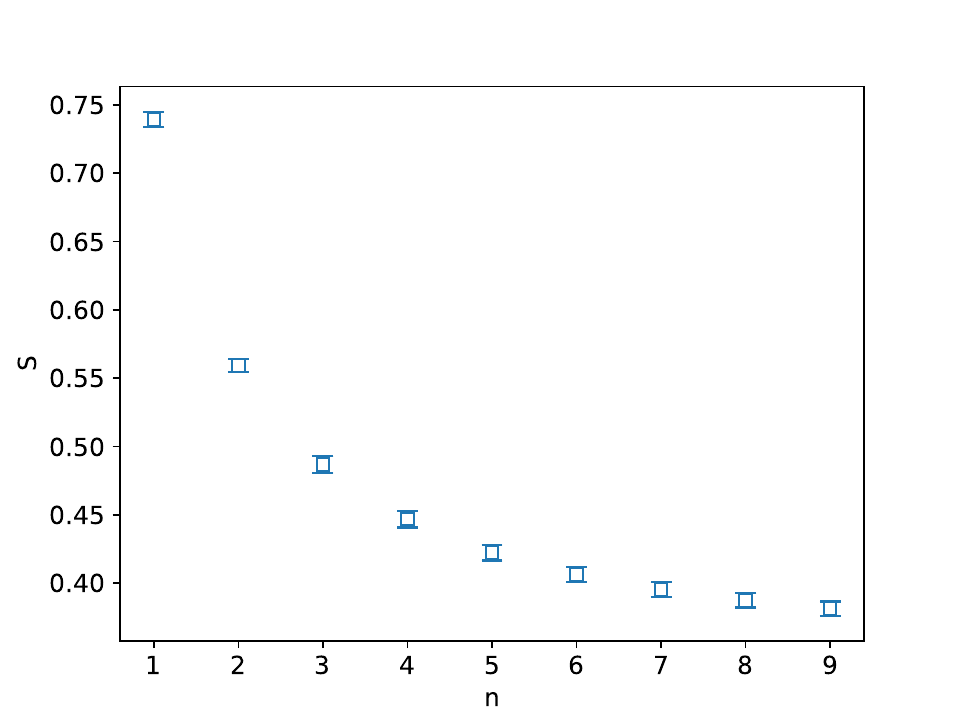}
    \caption{ Rényi entropies of order $n=2,3,\ldots 9$ compared with von Neumann entropy, plotted at $n=1$. Result was obtained for spin chain with $L=32$ spins at zero temperature, where the size of the subsystem $A$ is $l=5$ spins.}
    \label{fig:vN_vs_Ren}
\end{figure}

In Figure \ref{fig:vN_vs_Ren} we show a comparison of the Rényi entropies $S_n$ for $n=2,3,\ldots 8$ and the von Neumann entropy $S$ (at $n=1$). The values were obtained for $l=5$. By looking only at $n\ge 2$ values we have not found a simple form of the function which reproduces the value at $n=1$.

\section{Conclusions and outlook}

In this paper, we presented a method of calculating the reduced density matrix for spin systems and demonstrated it using the 1D quantum Ising model with $L=32$ spins. The method uses the path integral formalism to relate the 1D quantum model to a 2D classical model, which is then simulated using Monte Carlo methods based on autoregressive neural networks and the Neural Importance Sampling algorithm. The key feature of autoregressive networks is explicit access to the conditional probabilities of the spins, which allows us to use a single hierarchy of neural networks for the calculation of all density matrix elements at fixed time discretization and lattice volume.

The reduced density matrices are obtained for sizes of the subsystem equal to $l=1,2,3,4,5$. Those matrices were later diagonalized, and the eigenvalues were used to calculate von Neumann and Rényi entanglement entropies. Extrapolations allowed us to obtain results for entropies at zero temperature and in the continuum limit, where numerous conformal field theory (CFT) predictions are available. We found excellent agreement within our uncertainties for the von Neumann entanglement entropy. For the Rényi entanglement entropy, we fitted the dependence on $l$ predicted by CFT; however, we observed significant final size effects for small values of $l$.

Our method is not limited to zero temperature and can be applied to thermal states as well. Extraction of the entanglement entropies at some given temperature requires appropriate tuning of the size of the temporal extent $m$ as the lattice spacing $\Dt$ is decreased; taking the additional limit $T \to 0$ is not required. 

There are other future directions that may be explored in using our approach. For example, it is relatively straightforward to explore different Hamiltonians (also including defects) and boundary conditions of spin systems. Such studies are interesting from the CFT point of view \cite{2009JPhA...42X4009A,2022PhRvL.128i0603R}. Going beyond 1D quantum (2D classical) systems is also possible - the first generalization of hierarchical autoregressive networks to 3D classical systems is available \cite{Bialas:2025hxu}. The most important bottleneck for developing our method is its scaling with the system size $L$, hence, further development of neural architectures is necessary.

\section*{Acknowledgements}

We gratefully acknowledge the Polish high-performance computing infrastructure PLGrid (HPC Center: ACK Cyfronet AGH) for providing computer facilities and support within computational grants no. PLG/2024/017341, no. PLG/2023/016656. 
T.S. and D.Z. acknowledge the support of the Polish National Science Center (NCN) Grant No. 2021/43/D/ST2/03375. P.K. acknowledges the support of the Polish National Science Center (NCN) grant No. 2022/46/E/ST2/00346. D.Z. acknowledges the support of the Research Support Module under the program Excellence Initiative - Research University at the Jagiellonian University. This research was partially funded by the Priority Research Area Digiworld under the program Excellence Initiative – Research University at the Jagiellonian University in Kraków.

 \appendix

\section{Neural network details}
\label{app}

The HAN algorithm can, in principle, be implemented using various architectures of neural networks. Our implementation is based on the dense networks with some connections masked (weights multiplied by $0$). We used two masked linear layers, the first one with PreLU activation, and the second one with Sigmoid.

The networks were trained in four stages. In all of them, the batch size was set to 2048, and the learning rate was 0.003, 0.001, 0.0001, and 0.00001 with Adam optimizer. Each stage lasted 30000 epochs. Depending on the size of the sample, the whole training for one set of parameters lasted up to three hours (on Nvidia A100 GPU).

The efective sample size ($\ess$) is defined as:a
\begin{equation}
    \ess = \frac{\avg{\hat{w}}_{q_\theta}^2}{\avg{\hat{w}^2}_{q_\theta}} \approx \frac{\left(\sum_{i=1}^{\cal N} \hat w(\mathbf{s}_i)\right)^2}{{\cal N} \sum_{i=1}^{\cal N} \hat w^2(\mathbf{s}_i)}.
\label{ESS_definition}
\end{equation}
The $\ess \in (0,1]$ is an indicator of how well the networks were trained \cite{Liu} - the  higher $\ess$, the closer $q_\theta$ is to $p$. 

The ESS of the samples generated by the trained networks was in the range 0.0001 - 0.5, depending mainly on $k$, but also on $\Dt$. Small values of $\ess$ imply large variance of the weights $\hat{w}_A(\v s)$, but we have gathered enough   statistics to keep the errors small.  

For each density matrix element, there were 16 - 160 mln samples generated. The generation of 8 mln samples with batch size set to $2^{14}$ lasted from 90 s to 300 s depending on $k$.

\bibliographystyle{ieeetr}
\bibliography{references2}

\end{document}